\documentclass[runningheads]{llncs}
\usepackage[T1]{fontenc}
\usepackage{graphicx}

\usepackage[english]{babel}
\usepackage{array}

\usepackage{color}
\PassOptionsToPackage{hyphens}{url}
\usepackage[colorlinks=true,linkcolor=blue,citecolor=blue,urlcolor=blue]{hyperref}

\usepackage{orcidlink}
\usepackage{marvosym}
\newcommand{\orcidicon}[1]{\href{https://orcid.org/#1}{\raisebox{-1.29bp}{\resizebox{8.5bp}{8.5bp}{\orcidlogo}}}}
\renewcommand\orcidID[1]{\hspace{0.5bp}\orcidicon{#1}\hspace{0.49bp}}
\newcommand{\corrauthor}{\textsuperscript{(\raisebox{-0.42bp}{\fontsize{10.11bp}{12bp}\selectfont\Letter}\hspace{-1.49bp})}}
\makeatletter\let\llncs@makefntext\@makefntext\makeatother
\usepackage[bottom]{footmisc}
\makeatletter\let\@makefntext\llncs@makefntext\makeatother

\makeatletter
\def\ps@headings{\let\@mkboth\@gobbletwo
   \let\@oddfoot\@empty\let\@evenfoot\@empty
   \def\@evenhead{\normalfont\small\hspace{\headlineindent}\leftmark\hfil}
   \def\@oddhead{\normalfont\small\hfil\rightmark\hspace{\headlineindent}}
   \def\chaptermark##1{}%
   \def\sectionmark##1{}%
   \def\subsectionmark##1{}}
\def\ps@titlepage{\let\@mkboth\@gobbletwo
   \let\@oddfoot\@empty\let\@evenfoot\@empty
   \let\@evenhead\@empty\let\@oddhead\@empty}
\AtBeginDocument{\def\lastandname{\unskip,\newline and}}
\usepackage{etoolbox}
\patchcmd{\@maketitle}{\vskip .8cm}{\vskip .8cm\vskip 2.87bp}{}{}
\makeatother

\makeatletter
\AtBeginDocument{\let\llncs@doi\doi\renewcommand{\doi}[1]{\bib@doi#1\@nil}}
\def\bib@doi{\@ifnextchar\rawlink{\bib@raw}{\bib@std}}
\def\bib@raw\rawlink#1\@nil{#1}
\def\bib@std#1\@nil{\llncs@doi{#1}}
\makeatother

\begin{document}
\title{ASA: A Simulation Environment\texorpdfstring{\\}{ } for Evaluating Military Operational\texorpdfstring{\\}{ } Scenarios}
\titlerunning{ASA: A Simulation Environment for Evaluating Military}
\author{
Joao P. A. Dantas\corrauthor\orcidID{0000-0003-0300-8027} \and
Andre N. Costa\orcidID{0000-0002-2309-9248} \and
Vitor C. F. Gomes\orcidID{0000-0003-3239-2160} \and
\newline Andre R. Kuroswiski\orcidID{0000-0003-1549-2434} \and
Felipe L. L. Medeiros\orcidID{0000-0001-8796-8230} \and
Diego Geraldo\orcidID{0000-0003-1389-9142}\and
\newline Adrisson R. Samersla\orcidID{0009-0005-4559-1849}\and
Samara R. Silva\orcidID{0009-0000-5065-7236}\and
Andre F. M. Caetano\orcidID{0000-0001-5821-7729}\and
\newline Yuri D. Ferreira\orcidID{0009-0003-2583-3076}\and
Alessandro O. Arantes\orcidID{0000-0002-9024-1902}\and
Davison S. Santos\orcidID{0009-0004-3563-4473}\and
Marcia R. C. Aquino\orcidID{0009-0006-5118-5764}
}

\authorrunning{J. P. A. Dantas et al.}
\institute{Decision Support Systems Subdivision, Institute for Advanced Studies,\\
Trevo Coronel Aviador Jose A. A. do Amarante, 1, Putim, Sao Jose dos Campos,\\
12228-001 Sao Paulo, Brazil\\
\email{\{dantasjpad,negraoanc,vitorvcfg,kuroswiskiark,felipefllm,diegodg,}\\
\email{samerslaars,samarasrs,andreafmc,yuriydf,alessandroaoa1,}\\
\email{davisondss,marciamrca\}@fab.mil.br}}
\maketitle
\vspace*{-5.07bp}
\begin{abstract}
The Aerospace Simulation Environment (\emph{Ambiente de Simulação Aeroespacial -- ASA} in Portuguese) is a custom-made object-oriented simulation framework developed mainly in C++ that enables the modeling and simulation of military operational scenarios to support the development of tactics and procedures in the aerospace context for the Brazilian Air Force. This work describes the ASA framework and its main features: a distributed architecture for coordinating multiple simulation machines, a modular structure that allows models to be loaded at runtime, a batch execution mode for simulating multiple scenarios with varied initial conditions, and an integrated data analysis platform for post-processing simulation results. In addition, we present a list of recent studies that have used ASA in applications related to decision support and autonomy in air combat scenarios.

\par\vspace{5.72bp}\vspace{0pt}
\raggedright

\par\vspace{-3.75bp}\vspace{0pt}
\end{abstract}

\section{Introduction}
\vspace{-1.86bp}
\label{sec:introduction}

The Institute for Advanced Studies (IEAv), a research organization of the Brazilian Air Force (\emph{Força Aérea Brasileira} — FAB, in Portuguese), has developed, since 2018, the Aerospace Simulation Environment (\emph{Ambiente de\linebreak Simulação Aeroespacial} — ASA, in Portuguese) to provide a computational\linebreak solution that enables the modeling and simulation of operational scenarios.\linebreak This solution allows users to define strategies, parameters, and command\linebreak decisions to support the development of tactics, techniques, and procedures in the aerospace domain for defense purposes.

The characteristics of modern battlefield scenarios present significant challenges to the development of practical combat simulations~\cite{costa2025simulation}. These challenges call for more integrated and flexible solutions that can address both technical and organizational aspects~\cite{hill2018open}.

To meet some of these demands, frameworks such as the Advanced Framework for Simulation, Integration, and Modeling (AFSIM)~\cite{clive2015advanced}, Wukong~\cite{han2022deep}, FLAMES~\cite{flames}, and VR-Forces~\cite{vrforces} have been developed. However, many of\linebreak these tools are commercial products or restricted to use within specific countries. In this context, the ASA environment was conceived as a national solution designed to support FAB’s strategic planning, meet operational analysis needs, and promote the development and evaluation of emerging technologies for military research.

ASA is designed as a flexible and modular platform, capable of adapting to diverse user needs. This flexibility is essential given the wide range of requirements from its users. These needs could not be fully addressed by commercial off-the-shelf (COTS) simulation software. Rather than building an entirely new system, the development approach integrated openly available tools into a unified simulation environment that is flexible, accessible, and scalable.

The proposed solution uses the Mixed Reality Simulation Platform (MIXR)~\cite{hodson2018mixed} as its simulation engine, an open-source software project designed to support the development of robust, scalable, virtual, constructive, standalone, and distributed simulation applications. ASA extends MIXR’s capabilities by adding components that streamline tasks for both developers and analysts. A manager application was created to serve as an interface between multiple resources, working as a hub to execute, store, and analyze simulations across multiple machines. This application also supports the simultaneous creation of numerous simulations by varying initial conditions according to the analyst's needs. Furthermore, models and tools can be dynamically loaded at runtime to increase flexibility. All simulation data are stored in a dedicated database, expediting data collection and enabling more robust statistical analysis. Additionally, given the complexity of simulation outcomes and the varied technical background of ASA users, a dedicated data analysis platform was integrated into the system, not only for planning and visualization but also for post-processing the scenario data.

The main contribution of this work is the introduction of a new environment for modeling and simulation in the aerospace domain for military applications. It features: (i) a distributed architecture for managing multiple simulation\linebreak machines; (ii) support for runtime model loading within a modular architecture, allowing new models to be easily integrated; (iii) a batch execution mode that enables the simulation of multiple scenarios with varied initial conditions; and (iv) an enhanced data analysis platform for post-processing military operational scenario data. Additionally, we present a list of recent studies that have used the ASA platform in applications related to decision support and autonomy in military operational scenarios.

The remainder of this paper is organized as follows. Section~\ref{sec:architecture} presents the ASA architecture. Section~\ref{sec:applications} discusses studies that have used ASA in air combat analysis as application examples of the simulation framework. Finally, Sect.~\ref{sec:conclusion} provides concluding remarks on the current state of ASA and outlines potential directions for future work.

\section{ASA Architecture}
\label{sec:architecture}

The ASA design consists of three main modules. The first part is the simulation framework, defined as AsaSimulation, which provides the applications and necessary services to create and execute simulations. The second part comprises the interface applications, denominated AsaInterfaces, which provide tools for creating scenarios by listing all available components to be included and a library for interacting with the AsaSimulation module. Lastly, the third part is the analysis module, called AsaDataScience, which allows for post-processing and analysis of scenario executions.

\begin{figure}[ht]
\centering
\includegraphics[width=0.989\textwidth,trim={0 14.47bp 0 15.4bp}]{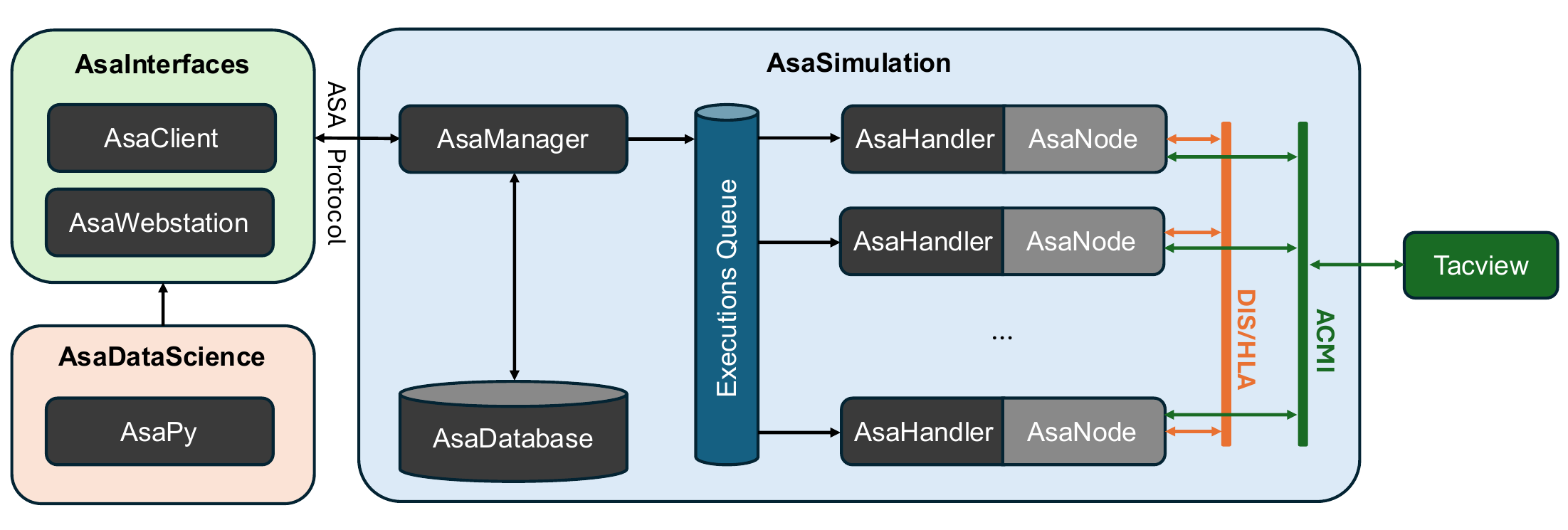}
\caption{ASA: modules and software applications.}\label{fig:architecture}
\end{figure}

Figure~\ref{fig:architecture} displays a summary of the ASA architecture, and the following subsections provide details on all three primary ASA modules. All ASA applications\linebreak use network communications, allowing processing to be distributed across multiple servers on a network.

\subsection{AsaSimulation}
\label{subsec:asasimulation}

The AsaSimulation module provides the necessary components for developing and executing a scenario simulation. It consists of applications, services, and libraries that help create agent models by developers and in the elaboration and simulation of scenarios by analysts.

The main features of this module are management and execution of simulations, dynamic loading of extension models, distribution of simulation executions, and management of simulation batch processing.

A permanent storage service called AsaDatabase keeps agent metadata, simulation scenarios, execution data, and analysis results. This service meets the storage demands of the AsaSimulation and AsaDataScience modules.

New agent models can be added to ASA by extending the interfaces and classes available in a library called AsaExtension. A new agent model that will be loaded into the AsaSimulation framework must provide its compiled source code (shared object file) and a JavaScript Object Notation (JSON) file that describes the parameters and components accepted by the model. AsaExtension also provides functionalities that allow extensions to store agent data in the AsaDatabase. This is done through the C++ macro \texttt{RECORD\_ASA\_CUSTOM\_DATA("tag", agent)}, which receives as its first parameter a tag that identifies the agent's type and the agent object with the attributes to be stored in the AsaDatabase. Storing the state of the agents at each simulation step is optional, but it is essential if there is an interest in performing post-processing of the simulation data.

The AsaSimulation framework enables analysts to specify agent parameters (placeholders) during the simulation execution request, facilitating the execution of batch simulations. Analysts can request a batch execution from these simulation scenario templates by providing a list of initial conditions for the previously selected agent parameters. Each set of initial conditions, combined with the scenario template, generates an execution request that is allocated to run on a distributed processing node.

The distribution of simulation processing is done by dividing the responsibilities of simulating to three applications on a network: AsaManager, AsaHandler, and AsaNode. The following subsections detail each of these applications.

\subsubsection{AsaManager} \label{subsec:asamanager}
is responsible for coordinating the processes to use ASA in a distributed manner.
One of the essential functions of AsaManager is receiving and preprocessing execution requests, and automatically and transparently dispatching the requested simulation to be executed on an available node.
The distribution of executions is performed using a queue service based on the Advanced Message Queuing Protocol (AMQP). AsaManager places a validated simulation execution request in the queue (Executions Queue) and, when available, an AsaHandler starts serving the request.

The use of queues to decouple direct management between request handling and execution enables the dynamic addition of new processing nodes, such as AsaHandler/AsaNode, to the system in response to increased demand for parallel execution. Because communication between AsaManager, ExecutionsQueue, and AsaHandler occurs over a network, applications can be distributed across different machines within a networked environment.

AsaManager incorporates a Representational State Transfer API (REST API) to enable interaction between other modules and the simulation functionalities, as well as access to stored data. This API implements the ASA protocol, which exposes AsaSimulation as a simulation service to other applications. It enables the submission, validation, execution, pausing, resumption, cancellation, and monitoring of single- or batch-simulation tasks. Additionally, this protocol defines methods for accessing data from completed simulations and managing user data.

Access is restricted to authenticated and authorized users who are permitted to use the system's functionalities. For each simulation execution request, AsaManager verifies the validity of the scenario to be simulated and checks whether the authenticated user is authorized to use the listed components and has sufficient credit available. Within the ASA framework, these credits are referred to as AsaCoins, which serve as a means to account for user resource consumption. During each simulation execution, both the CPU time consumed and the disk space used to store the simulated agent data were measured. These two variables are then converted into an equivalent value in AsaCoins, which is subsequently debited from the user's account. The reference rate for a single AsaCoin unit was established using the execution of a standard benchmark scenario. Based on this reference, the CPU time conversion rate is 0.0137650 AsaCoin per second, and the storage conversion rate is 0.0141355 AsaCoin per kilobyte.

\subsubsection{AsaHandler} \label{sec:asahandler}
is responsible for monitoring the execution queue, converting requests, and supervising simulation execution. When an AsaHandler is available and a new execution request is placed in the queue by AsaManager, the AsaHandler retrieves the request, converts it into a format compatible with the AsaNode, initiates a new AsaNode process, and monitors its execution. Communication between AsaHandler and AsaNode is carried out via interprocess communication, enabling AsaHandler to oversee the simulation run. If the AsaNode terminates unexpectedly or a failure is detected by the AsaHandler, it will terminate the execution and will keep the AsaManager informed of the simulation status.

\subsubsection{AsaNode}
\label{sec:asanode}
is the simulation engine of the ASA platform, and its primary function is to process the simulation itself. It is an executable file obtained from the compilation of codes developed in MIXR and features developed by the ASA team, such as dynamically loading extensions and controlling the simulation's execution (pause, resume, stop, execution speed, etc.). It estimates how the scenario will evolve, considering the models incorporated in each agent present in the simulation. AsaNode can run on the same machine as the AsaManager application or in a clustered computing environment. This capability is essential when the user wants to simulate a set of scenarios, called batch, and the main difference between them is the initial configuration of each agent.

At predefined time intervals, AsaNode reports the progress of the simulation to AsaHandler, allowing this application, or AsaManager, to manage the execution process. If enabled by the execution requester, AsaNode can transmit data over the network using either the Distributed Interactive Simulation (DIS) protocol or the High-Level Architecture (HLA) standard, both of which are supported by the MIXR framework. DIS is a protocol designed for real-time exchange of information between simulation entities in a distributed environment, while HLA is a general architecture that enables interoperability among different simulation systems through a shared runtime infrastructure~\cite{kuhl1999creating}.
To allow simulation data to be streamed for visualization in the Tacview flight analysis tool\footnote{\url{https://www.tacview.net}.}, the AsaExtensions library implements the ACMI file format and its associated Real-time Telemetry Protocol~\cite{tacview}.

\subsection{AsaInterfaces}
\label{sec:asainterfaces}

The AsaInterface module includes two tools: WebStation and AsaClient. WebStation provides a visual interface for building simulation scenarios, while AsaClient manages requests to the simulation manager and handles user authentication.

\subsubsection{WebStation} \label{sec:webstation}
provides a graphical user interface (GUI) accessible to registered users through a web page developed with VueJS and Django. Analysts use the platform to construct simulation scenarios that include military symbols, geometric drawings, aeronautical charts, and digital terrain models. The interface displays how each scenario progresses during the simulation, offering an interactive environment for analysis and planning. Users can define performance metrics and examine the results after the simulation is completed. The platform also allows users to specify which messages should be stored in the database during execution. Furthermore, WebStation includes a validation feature that checks whether scenario components meet the model’s requirements and ensures that all attributes and subcomponents are properly defined.

To support batch execution, WebStation allows users to configure input parameters that vary across multiple runs. After completing the scenario setup, users can download a JSON file containing the scenario data or save it to the database for future access. The interface also supports uploading JSON files, enabling users to edit and reuse existing scenarios.

\begin{figure}[ht]
\centering
\vspace*{0.37bp}
\includegraphics[width=12cm]{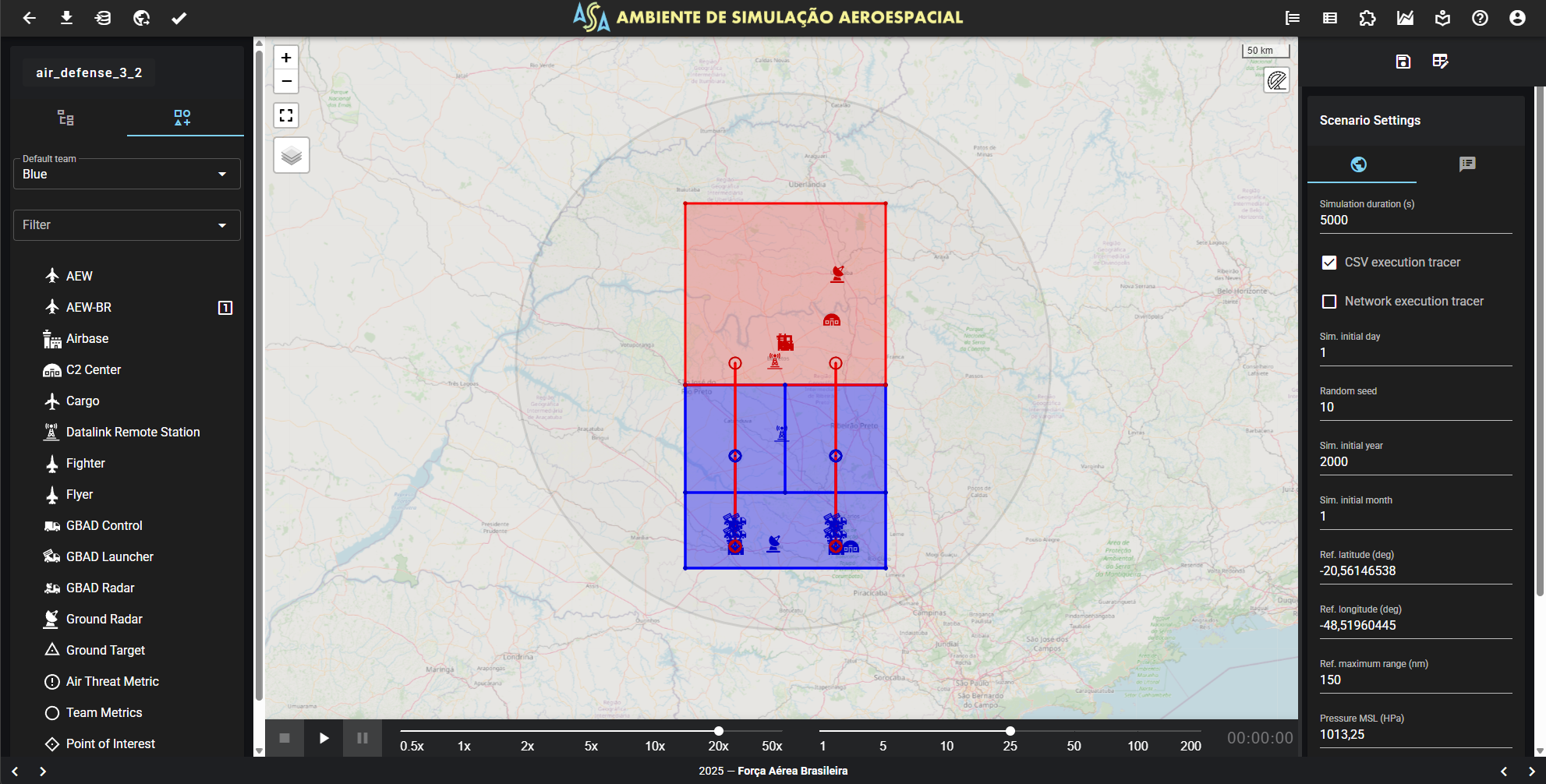}
\vspace*{1.1bp}
\caption{WebStation: interface to create and visualize simulations.}
\label{fig:webstation}
\end{figure}

An overview of the WebStation interface, including scenario construction and visualization features, is shown in Fig.~\ref{fig:webstation}.

\subsubsection{AsaClient} \label{sec:asaclient}
is a Python library designed to operate with the ASA protocol. It performs two primary functions: managing user authentication and handling communication with the simulation manager. AsaClient serves as an essential interface between the AsaDataScience environment and the Manager, enabling secure and structured access to simulation resources.

The library provides a set of high-level methods that allow analysts to interact with the core components of the simulation infrastructure. Through AsaClient, users can authenticate in the system, submit execution requests, and monitor simulation progress. AsaClient also supports database access for retrieving results, storing output, and managing simulation metadata.

\subsection{AsaDataScience}
\label{sec:asadatascience}

The AsaDataScience module provides tools for understanding the factors that led to the simulation results, supporting analysts in identifying how performance, cost, and operational constraints evolve across military scenarios. It is composed of two main components: (i) the AsaPy Python library, designed to assist in the post-processing of simulation data; and (ii) a cloud-based computational environment with access to high-performance hardware and a JupyterHub interface for large-scale batch execution and interactive data analysis.

At the core of this module is AsaPy, a specialized Python library that provides a structured pipeline for simulation data analysis~\cite{dantas2024asapy}. It integrates well-established methods for experiment planning, batch execution control, statistical analysis, and machine learning, enabling users to derive actionable insights with minimal programming effort. Its internal structure follows a typical analyst workflow and is composed of four main functional blocks, described in Table~\ref{tab:asapy}.

Running a simulation invariably results in the collection of massive data at every step. Once these data are adequately stored and structured in the AsaDatabase, analysts can employ techniques such as data visualization, statistical testing, and model fitting to uncover insights from the simulations. AsaPy streamlines this process by automating routine steps and standardizing the analytical workflow.

\begin{figure}[ht]
    \centering
    \includegraphics[width=0.9914\textwidth,trim={0 14.4bp 0 3.55bp}]{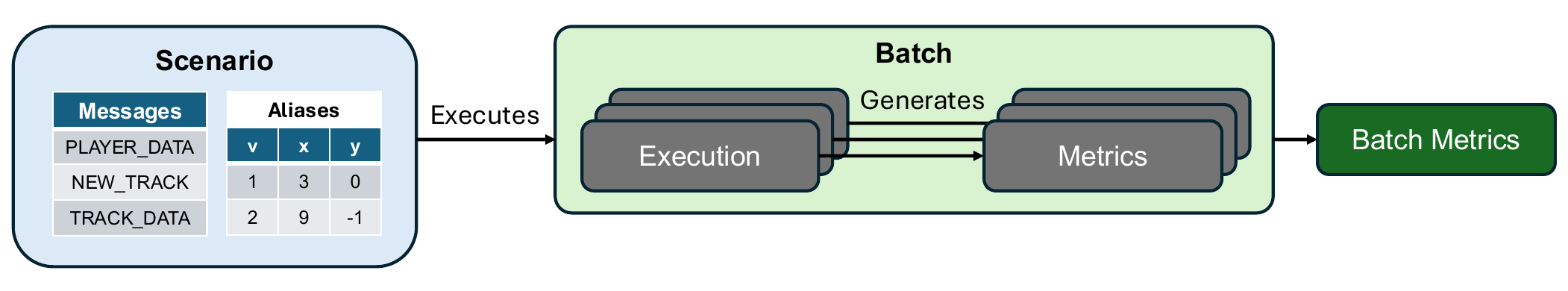}
    \caption{Batch execution flow supported by AsaPy. The process begins with the selection of aliases, which define the input parameters to be varied across simulations, and messages, which specify the simulation outputs to be collected. The scenario batch is then divided into sequential chunks. Each chunk is executed, and the selected outputs are used to compute batch-level metrics, which can support monitoring and early stopping of the execution process.}
    \label{fig:batch}
\end{figure}

A particularly important feature of AsaPy is its support for batch execution. In many military simulations, it is necessary to run hundreds or thousands of scenarios with varying input parameters~\cite{gill2018regression}. AsaPy allows these executions to be divided into chunks, monitors the results of each chunk, and applies convergence-based early stopping criteria to avoid unnecessary runs. This improves both analytical efficiency and resource usage. Figure~\ref{fig:batch} illustrates this batch execution workflow.

% Table styled and placed as in the published version
\newcommand{\tabrule}[1]{\noalign{{\color[gray]{0.5}\hrule height #1}}}
\begin{table}[t]
\caption{AsaPy functionalities and corresponding descriptions.}
\label{tab:asapy}
\small
\centering
\renewcommand{\arraystretch}{1.22}
\setlength{\tabcolsep}{0.6bp}
\begin{tabular}{>{\raggedright\arraybackslash}p{91.23bp}!{\color[gray]{0.5}\vrule width 0.4bp}>{\raggedright\arraybackslash}p{249.07bp}}
\tabrule{0.5bp}
Functionality & Description \\
\tabrule{1bp}
Design of Experiments & Generates input configurations to explore the simulation parameter space efficiently. \\
\tabrule{0.4bp}
Execution Control & Manages simulation batches with chunking and early stopping based on convergence of key metrics. \\
\tabrule{0.4bp}
Analysis & Offers tools for statistical analysis, data exploration, and visualization. \\
\tabrule{0.4bp}
Prediction & Applies supervised and unsupervised machine learning models for regression and classification tasks. \\
\tabrule{0.4bp}
\end{tabular}
\end{table}

To support large-scale simulations, AsaDataScience includes a cloud-based computational environment equipped with high-performance hardware and a JupyterHub interface\footnote{\url{https://jupyter.org/hub}.}. This setup enables analysts to manage simulation studies interactively through notebooks, providing both scripting flexibility and real-time visualization. JupyterHub also facilitates collaborative work by allowing multiple users to run batch executions, monitor progress, and analyze results in a shared and scalable environment, all without requiring direct access to low-level computing resources.

Overall, AsaDataScience serves as a unified environment for analyzing simulation data in defense applications, regardless of the simulation engine employed. In addition to supporting ASA-native data formats, it is also compatible with outputs from other frameworks, as demonstrated in \cite{dantas2022supervised}.

\vspace{-0.75bp}\vspace{0pt}
\section{Applications in Decision Support and Autonomy}
\vspace{0.45bp}
\label{sec:applications}

ASA has been a flexible and useful platform for research in air combat operations, especially in areas related to decision support and autonomous systems. The studies listed below, in chronological order, demonstrate how ASA has been gradually adopted in increasingly complex and realistic scenarios.

The earliest study, \cite{dantas2018apoio}, presented an artificial neural network model to support pilot situational awareness in Beyond Visual Range (BVR) air combat. Using simulation data, the system generated offensive and defensive assessments to aid decision-making during flight. This work was developed using the AEROGRAF Platform~\cite{Petersen2008}, which preceded ASA and influenced its initial design.

ASA was later used in \cite{kuroswiski2020modelagem} to explore the feasibility of agent-based modeling and simulation for assessing air defense capabilities in BVR engagements. This marked the first application of ASA in a strategic planning context.

In \cite{dantas2021weapon}, over 50{,}000 missile launches were simulated to train a deep neural network for estimating the maximum launch range of a missile's Weapon Engagement Zone (WEZ), achieving high predictive accuracy while reducing the need for further simulations. This study used the same missile model implemented in ASA.

In \cite{dantas2021engagement}, 3{,}729 BVR engagements were simulated to train a supervised learning model aimed at supporting the decision of when to engage enemy aircraft during Defensive Counter Air (DCA) missions. The model was trained using simulation-derived operational metrics to capture relevant combat dynamics.

In \cite{dantas2022machine}, ASA/AEROGRAF was used to simulate 10{,}000 BVR combat scenarios, enabling the development of classifiers that improved pilot situational awareness. The models achieved high accuracy in distinguishing between offensive and defensive conditions.

The work in \cite{costa2022formation} focused on aircraft formation control using artificial potential fields and optimization techniques. ASA simulations helped identify configurations that improved formation coherence and mission success.

The study \cite{limafilho2022optimization} employed ASA to optimize tactical unmanned aerial vehicle (UAV) formations under uncertainty in war-game scenarios. Metaheuristics were applied to maximize the probability of success against opposing forces.

In \cite{dantas2023autonomous}, ASA was used to propose an architecture for training a deep reinforcement learning agent capable of autonomously learning BVR tactics.

The work in \cite{dantas2023real} proposed an approach for analyzing surface-to-air missile engagement zones using prediction models to enable real-time estimation while reducing computational costs. The analysis was based on the same surface-to-air missile model implemented in ASA.

The work in \cite{dantas2024loyal} proposed new social navigation metrics to assess collaboration between human pilots and autonomous wingmen in air combat scenarios. It also outlined a validation experiment using ASA to simulate mixed human-autonomous formations.

In \cite{viscardi2024advanced}, a defensive ``winding maneuver'' was proposed to improve aircraft survivability against surface-to-air missiles. Developed using ASA, the maneuver was optimized with a genetic algorithm and validated through simulations, statistical analysis, and operational metrics.

The work in \cite{lima2024aprendizado} proposed machine learning models to approximate ASA simulation outcomes in stochastic contexts, significantly reducing overall simulation time.

The study in \cite{dantas2025poker} introduced a probabilistic kill estimation model for air-to-air missiles based on ASA simulations and stochastic target modeling. It employed the same missile structure used in ASA.

Finally, \cite{dantas2024autonomous,dantas2025autonomous} used pilot data from ASA/AEROGRAF to train imitation learning models for replicating pop-up attack maneuvers. A variational autoencoder was employed to generate synthetic samples and augment the dataset, improving the robustness and performance of the models.

These studies collectively demonstrate the flexibility and effectiveness of ASA as a simulation tool for operational analysis, autonomy development, and tactical decision support in air combat environments.

\vspace{-5.89bp}\vspace{0pt}
\section{Conclusion and Future Work}
\vspace{-2.94bp}
\label{sec:conclusion}

In this work, we presented a high-level overview of the ASA simulation framework, developed by IEAv since 2018, with the primary objective of supporting the evaluation of military operational scenarios relevant to FAB. The platform offers several distinguishing features, including the management of multiple simulation machines across one or more computers, as well as the dynamic loading \texttt{.so} files at runtime, batch execution of simulations, and the AsaDataScience module, an integrated data analysis platform tailored for military operational studies. Furthermore, we highlighted recent works that have employed ASA as a simulation tool to support air combat applications.

For future work, we plan to release part of the ASA source code, including its general architecture, to a selected group of organizations. This controlled release aims to encourage the development of diverse applications within the same simulation platform while ensuring proper control and security. Additionally, ASA is expected to evolve into a Simulation-as-a-Service (SimaaS) tool, supporting a wide range of simulation demands in the defense and aerospace domains~\cite{dantas2023asasimaas}. This initiative aims to promote greater interoperability among government, academia, and industry.

From a technical perspective, several enhancements are under consideration. First, we aim to enhance scalability through parallel processing by enabling the flexible addition of simulation nodes as computational demands increase. Another area of focus is observability: ASA will incorporate mechanisms to\linebreak monitor resource usage (such as CPU, memory, and disk), track feature utilization (identifying which components and libraries are used most frequently), and detect failures. These capabilities will enable better instrumentation, increased failure tolerance, and enhanced platform operational visibility.

We also aim to introduce accountability mechanisms to estimate and control the computational cost of each simulation, allowing resource usage to be better quantified and managed. Regarding the core simulation engine, currently based on an object-oriented paradigm with deep inheritance hierarchies, we are exploring a transition to a more modular and maintainable design. One promising alternative is the adoption of the entity-component-system (ECS) paradigm, which can improve model readability and simplify the integration of new components~\cite{muratet2020accessibility}.

Additionally, improvements are planned for the internal structure of the \mbox{AsaPy} library, with efforts to reorganize its modules for better maintainability and extensibility. Finally, we expect to adapt the framework to support more modular and flexible decision-making processes in autonomous agent models by developing an artificial intelligence server in Python that integrates a rich ecosystem of specialized tools and libraries.

\begin{credits}
\subsubsection{\ackname} This work has been supported by the Funding Authority for Studies and Projects (FINEP)---Finance Code 01.20.0196.00/2824/20.

\vspace{-0.5bp}\vspace{0pt}
\subsubsection{\discintname}
The authors have no competing interests to declare that are relevant to the content of this article.
\end{credits}

\bibliographystyle{splncs04}
\bibliography{references}
\end{document}